\documentclass[aps,prl,reprint,amsmath,amssymb]{revtex4-1}
\usepackage{graphicx}
\begin{document}

\title{Nonlinear absorption in dielectric metamaterials}

\author{Brian Slovick}
\altaffiliation{brian.slovick@sri.com}
\author{Lucas Zipp}
\author{Srini Krishnamurthy}
\affiliation{%
Applied Optics Laboratory, SRI International, Menlo Park, California 94025, United States}%

\date{\today}

\begin{abstract}
We solve the nonlinear Maxwell equations in an InP-based dielectric metamaterial, considering both two-photon absorption and photo-induced free-carrier absorption. We obtain the intensity-dependent reflection, absorption, and effective permittivity and permeability of the metamaterial. Our results show that nonlinear absorption dampens both the electric and magnetic Mie resonance, although the magnetic resonance is more affected because it occurs at longer wavelengths where the free-carrier absorption cross section is larger. Owing to field concentration in the metamaterial at resonance, the threshold intensity for nonlinear absorption is smaller by a factor of about 30 compared to a homogeneous layer of the same thickness. Our results have implications on the use of dielectric metamaterials for nonlinear applications such as frequency conversion and optical limiting.

\end{abstract}

\maketitle

Materials with strong nonlinear response are desired for applications involving optical limiting \cite{Stryland1985,Boggess1985,Stryland1988}, ultrafast modulation \cite{Scalora1994,Tran1997,Hache2000}, frequency conversion \cite{Rashkeev2001,Lekse2009}, and optical isolation \cite{Gallo1999,Yu2009}. Traditional bulk materials have a weak interaction between electrons and sub-bandgap photons, as measured by nonlinear absorption coefficients \cite{Chang2014,Bechtel1976,Boggess1986,Krishnamurthy2011}, and thus require high intensities or long interaction lengths to achieve an efficient nonlinear response.

One way to enhance the nonlinear response is to incorporate metamaterial elements or plasmonic structures to concentrate the electric fields within a nonlinear material \cite{Kauranen2012,Schuller2010}. Metamaterials containing plasmonic nanostructures have been integrated with nonlinear materials to enhance second-harmonic generation \cite{Czaplicki2013,Aouani2012,Thyagarajan2012,Thyagarajan2013,Zhang2011,Harutyunyan2012,Navarro2012} and achieve analog electromagnetically induced transparency \cite{Gu2012,Kurter2011}. The drawback of plasmonic approaches is that the fields are highly localized at the metal dielectric interface, leading to small interaction volumes \cite{Shcherbakov2014,Yang2015}. Also, the finite conductivity of metals at optical frequencies leads to undesirable losses. An alternative approach is to induce nonlinearity in dielectric metamaterials using Mie resonances \cite{Kapitanova2017}. Dielectric nanostructures have been used to enhance third-harmonic generation \cite{Shcherbakov2014,Yang2015,Shcherbakov2015a,Smirnova2016,Grinblat2016,Shorokhov2016} and to achieve ultrafast optical modulation \cite{Shcherbakov2015b,Makarov2015,Baranov2016}.

To date, the research on nonlinear phenomena in dielectric metamaterials has focused primarily on the experimental aspects \cite{Shcherbakov2014,Yang2015,Shcherbakov2015a,Smirnova2016,Grinblat2016,Shorokhov2016}, with relatively few examples of theoretical studies. A few examples employ the recently-developed linear generalized source method for nonlinear materials \cite{Weismann2015,Weismann2016,Kruk2015}, which calculates the diffraction of one- and two-dimensional gratings accounting for nonlinear polarization sources. Although these models provide important insights, they do not support three-dimensional structures and do not represent full solutions of the nonlinear Maxwell equations. Also, existing models do not account for the nontrivial frequency dependence of the nonlinear parameters, such as the two photon absorption (TPA) coefficient and the free-carrier absorption (FCA) cross section \cite{Krishnamurthy2011}.

In this work, we develop a full-wave model to solve the nonlinear Maxwell equations in a structured, three-dimensional metamaterial accounting for both TPA and photo-induced FCA. The nonlinear absorption coefficients are obtained from full-band structure calculations \cite{Krishnamurthy2011}. We apply the model to study the optical properties and effective parameters of a representative indium phosphide (InP)-based dielectric metamaterial operating in the near infrared spectral band. As expected, we find that nonlinear absorption at high intensities leads to dampening of the electric and magnetic Mie resonances. For continuous wave illumination, the onset of nonlinear absorption occurs at intensities of 1 MW/cm$^2$, while the Mie resonances are almost completely diminished for intensities approaching 5 MW/cm$^2$. In addition, we find several unexpected results. First, the nonlinear absorption at the magnetic resonance is larger than at the electric resonance, which is explained by the wavelength-dependent FCA. Second, assuming FCA is independent of wavelength, the absorption at the two resonances is found to be nearly equal, despite the electric field being heavily localized at the electric resonance and more uniformly distributed at the magnetic resonance. Third, owing to the enhancement of the electric field at resonance, we find that the intensity threshold of nonlinear absorption in the metamaterial is nearly 30 times lower compared to a homogeneous material of similar thickness.

To clearly illustrate the effects of nonlinearity on Mie resonances, we consider a large index-contrast metamaterial consisting of 360 nm InP spheres with large refractive index (3.3 at 1 $\mu$m) in air medium, arranged in a square lattice with a periodicity of 800 nm, as shown in Fig. 1(a). InP is chosen for its transparency in the near-infrared band of interest (0.9-1.3 $\mu$m). Note that the conclusions drawn in this article are valid even if the air medium is replaced by a polymer and the metamaterial layer is placed on a low index substrate such as silica. The size and periodicity of the InP spheres are optimized to position the electric and magnetic Mie resonances, identified as narrowband peaks in the reflection spectrum, in the band of interest as shown in Fig. 1(b).

\begin{figure}
\includegraphics[width=64mm]{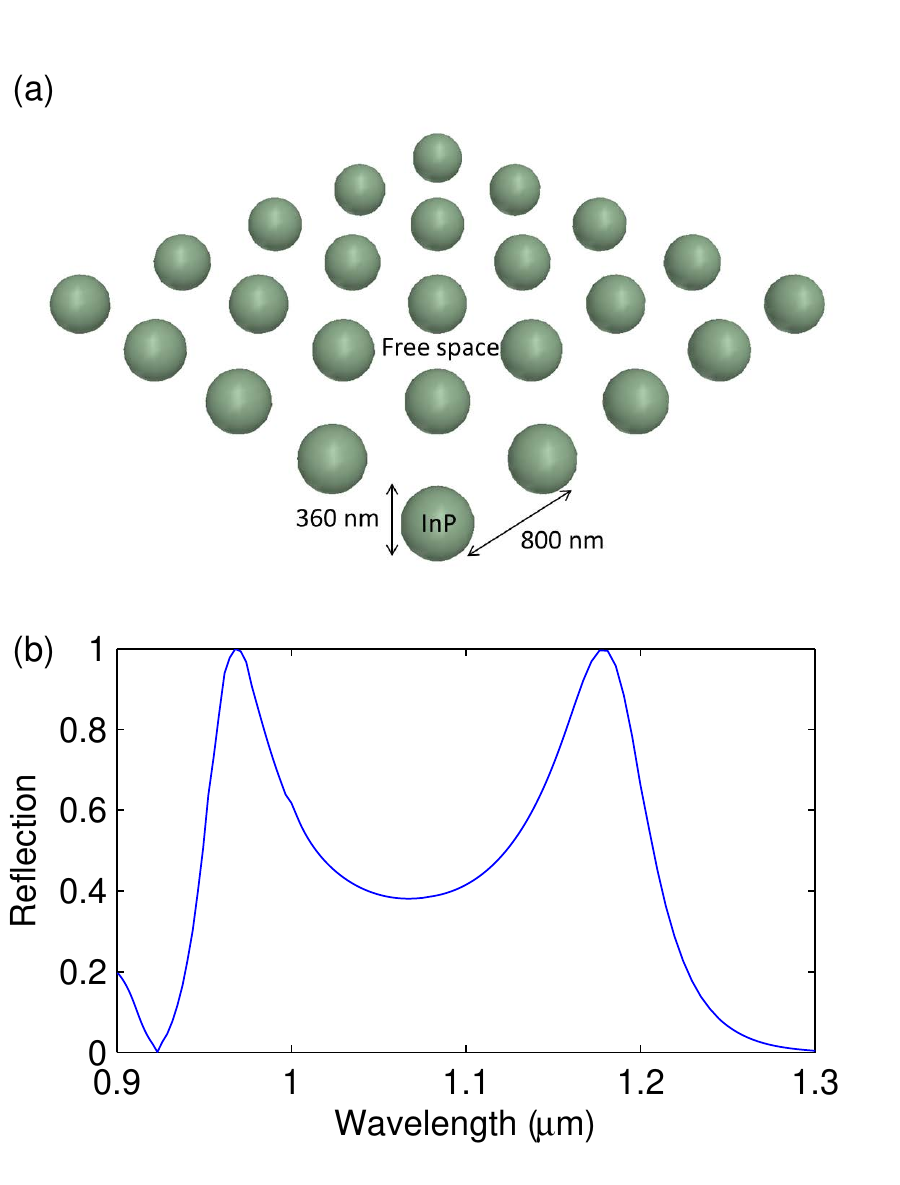}
\caption{\label{fig:epsart} (a) Dielectric metamaterial consisting of InP spheres in free space. (b) The calculated reflection spectrum for low intensities, showing two peaks corresponding to the electric and magnetic Mie resonances.}
\end{figure}

\begin{figure}
\includegraphics[width=30mm]{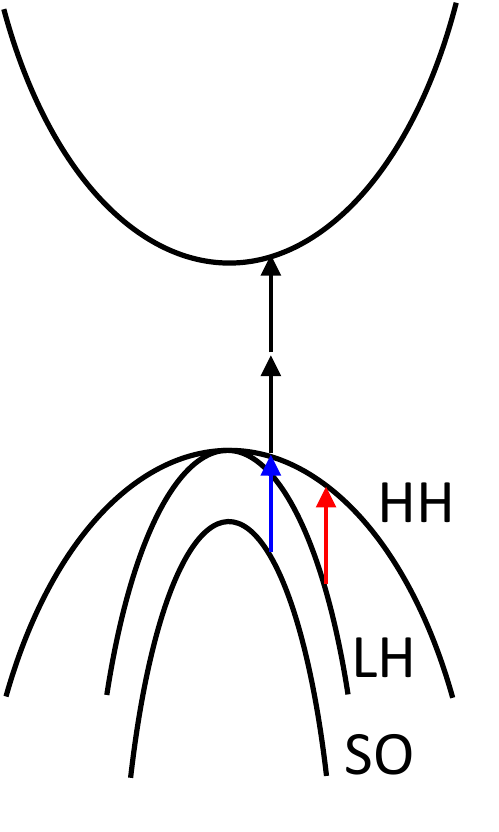}
\caption{\label{fig:epsart} Electronic band structure of InP, showing heavy hole (HH), light hole (LH), and spin orbit (SO) bands.}
\end{figure}

The origin of nonlinear absorption in InP can be understood from the band structure, shown in Fig. 2. The valence bands (VBs) consist of heavy-hole (HH), light-hole (LH), and spin-orbit (SO) bands. The conduction band (CB) is separated from the HH band by the band gap. In the absence of light, states in the VB are filled with electrons and the CB states are empty. Electrons in the VB can absorb photons with energy larger than the band gap and enter the CB. Since the band gap of InP (1.45 eV) is larger than the photon energies in the band of interest (0.95-1.4 eV), at low intensities photons transmit through InP without absorption. However, when the incident intensity is high, the probability for valence electrons to absorb two photons (shown as two stacked vertical arrows) is high, resulting in reduced transmission. In addition, this TPA is followed by FCA in which the holes left behind in the HH band can be filled by one-photon absorption by electrons in the LH and SO bands, shown by the colored arrows in Fig. 2. Because the strength of both TPA and FCA depend on intensity, they are referred to as nonlinear absorption processes. Figure 3 shows the previously calculated values of the TPA coefficient $\beta$ and the FCA cross section $\sigma$ for InP \cite{Krishnamurthy2011}. The value of $\beta$ is relatively constant with wavelength, which is typical for wide-bandgap materials, while $\sigma$ increases by an order of magnitude over the band. The FCA increases with increasing wavelength because the corresponding photon energy decreases, and the energy-momentum conservation condition for FCA (colored lines in Fig. 2) is satisfied only near the center of the Brillouin zone, where a larger number of holes are present.

\begin{figure}
\includegraphics[width=71mm]{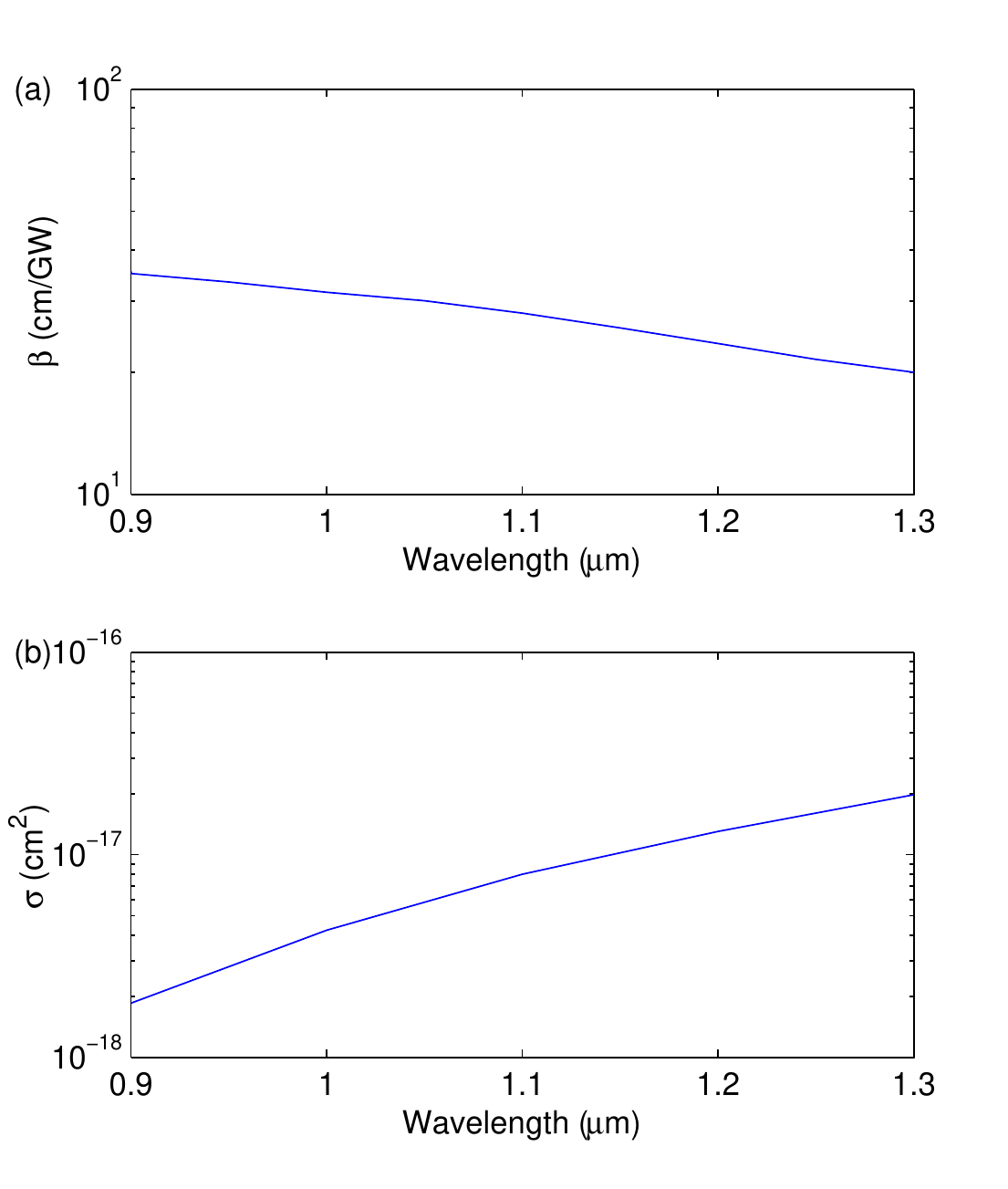}
\caption{\label{fig:epsart} Wavelength dependence of the two-photon absorption coefficient (a) and free-carrier absorption cross section (b) for InP \cite{Krishnamurthy2011}.}
\end{figure}

We will now incorporate these nonlinear coefficients into Maxwell's equations. The nonlinear Maxwell equation for the electric field $\textbf{E}(\textbf{r},t)$ is \cite{Shen1984,Stegeman2012}
\begin{equation}
\left(\nabla^2 -\frac{n^2}{c^2}\frac{\partial^2 }{\partial t^2}-\frac{n\sigma N}{c}\frac{\partial }{\partial t}\right) \textbf{E}(\textbf{r},t)=\frac{1}{\epsilon_0 c^2} \frac{\partial^2 \textbf{P}_{NL}(\mathbf{r},t)}{\partial t^2},
\end{equation}
where $n$ is the refractive index, $c$ is the speed of light, $N$ and $\sigma$ are the free-carrier concentration and absorption cross section, $\epsilon_0$ is the free-space permittivity, and $\textbf{P}_{NL}(\textbf{r},t)$ is the nonlinear polarization. Taking the Fourier transform of Eq. (1), assuming an $\exp{(-i\omega t)}$ time dependence, and using the relation for the third-order nonlinear polarization
\begin{equation}
\textbf{P}_{NL}(\textbf{r},\omega)=\frac{3}{4}\epsilon_0 \chi^{(3)}(\omega) |\textbf{E}(\textbf{r},\omega)|^2 \textbf{E}(\textbf{r},\omega),
\end{equation}
 where $\chi^{(3)}$ is the nonlinear susceptibility, we obtain
\begin{equation}
\left( \nabla^2 +n^2 \frac{\omega^2}{c^2}+\frac{3}{4} \frac{\omega^2}{c^2} \chi ^{(3)}|\textbf{E}(\textbf{r},\omega)|^2  +i\omega \frac{n\sigma N}{c} \right) \textbf{E}(\textbf{r},\omega)=0.
\end{equation}
Relating the imaginary part of $\chi^{(3)}$ to the two-photon absorption coefficient $\beta$ as \cite{Shen1984,Stegeman2012}
$$
\text{Im}(\chi^{(3)})=\frac{2n^2c^2\epsilon_0}{3 \omega} \beta,
$$
and neglecting the real part of $\chi^{(3)}$, Eq. (3) can be rewritten as
\begin{equation}
\left( \nabla^2 +n^2 \frac{\omega^2}{c^2}+i\frac{1}{2} \beta \omega n^2 \epsilon_0 |\textbf{E}(\textbf{r},\omega)|^2+i\omega \frac{n\sigma N}{c} \right) \textbf{E}(\textbf{r},\omega)=0.
\end{equation}
The free carrier concentration is given by the continuity equation for free electrons \cite{Shen1984,Krishnamurthy2011,Stegeman2012}
\begin{equation}
\frac{dN}{dt}=\frac{\beta I^2}{2 \hbar \omega}-\frac{N}{\tau},
\end{equation}
where $I=\frac{1}{2} nc\epsilon_0 |\textbf{E}(\textbf{r},\omega)|^2$ is the intensity, $\hbar \omega$ is the photon energy and $\tau$ is the photo-carrier relaxation time, which we assume is a constant equal to 1 $\mu$s. The first term in Eq. (5) describes free-carrier generation via TPA, and the second term describes free carrier recombination. For continuous-wave illumination, the free-carrier concentration will reach steady state conditions ($dN/dt=0$) and thus
\begin{equation}
N(I)=\frac{\beta \tau}{2\hbar \omega}I^2.
\end{equation}
Substituting Eq. (6) into Eq. (4), we obtain the following form of the nonlinear Maxwell equation:
\begin{eqnarray}
 \bigg( \bigg. \nabla^2 +n^2 \frac{\omega^2}{c^2}&&+i\frac{1}{2} \beta \omega n^2 \epsilon_0 |\textbf{E}(\textbf{r},\omega)|^2  \nonumber \\
&&+i \frac{\sigma \beta \tau n^3 c \epsilon_0^2}{8\hbar}|\textbf{E}(\textbf{r},\omega)|^4 \bigg. \bigg) \textbf{E}(\textbf{r},\omega)=0.
\end{eqnarray}
We solve Eq. (7) using the full-wave finite-element frequency domain solver in COMSOL. This was accomplished by assigning the two nonlinear terms in Eq. (7) to the imaginary part of $n^2$. Before solving Eq. (7) in the metamaterial in Fig. 1, we apply it to a homogeneous nonlinear medium and compare the results with the solution to the well-known rate equation \cite{Shen1984,Stegeman2012}
\begin{equation}
\frac{dI}{dx}=-\beta I^2-\frac{\sigma \beta \tau}{2\hbar \omega} I^3.
\end{equation}
The transmitted intensity, as a function of thickness for InP at a wavelength of 1 $\mu$m for various intensities, calculated by solving Eq. (7) and (8), respectively, are shown by dots and solid lines in Fig. 4. The two calculations are in excellent agreement, thus validating our full-wave nonlinear model. In this validation the index of InP is set equal to 1 to avoid interference effects, which are not included in Eq. (8).

\begin{figure}
\includegraphics[width=64mm]{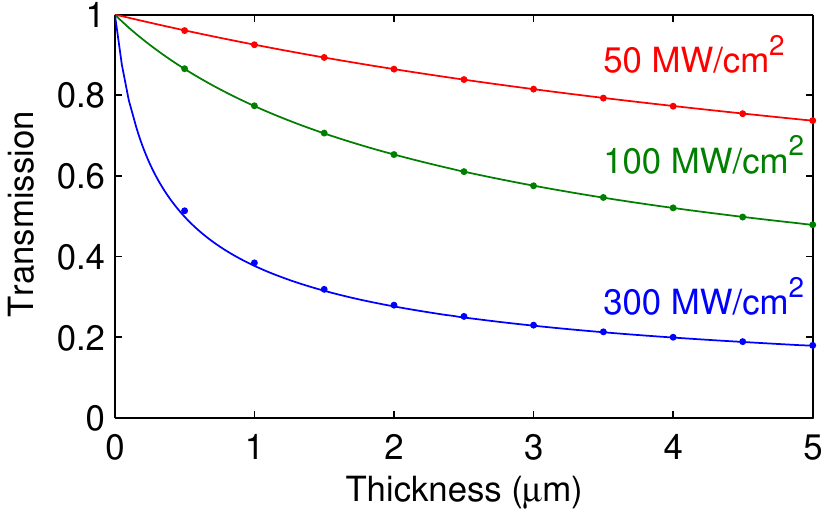}
\caption{\label{fig:epsart} Transmission of bulk InP at 1 $\mu$m as a function of thickness for different incident intensities, calculated using nonlinear full-wave COMSOL (dots) and by numerically integrating Eq. (8) (solid lines).}
\end{figure}

We now apply our validated nonlinear full-wave model [(Eq. (7)] to understand the role of nonlinear absorption in the dielectric metamaterial shown in Fig. 1(a). First, we studied the reflection and absorption for different incident intensities, shown in Fig. 5. For all intensities, the reflection spectrum contains narrowband peaks near 1.2 and 0.95 $\mu$m, corresponding to the magnetic and electric dipole Mie resonances, respectively. For a low intensity of 1 W/cm$^2$, the nonlinear processes are negligible, resulting in low absorption and nearly 100\% reflection at the two resonances. As the intensity increases, the absorption at the resonances increases and the reflection decreases. 

\begin{figure}
\includegraphics[width=65mm]{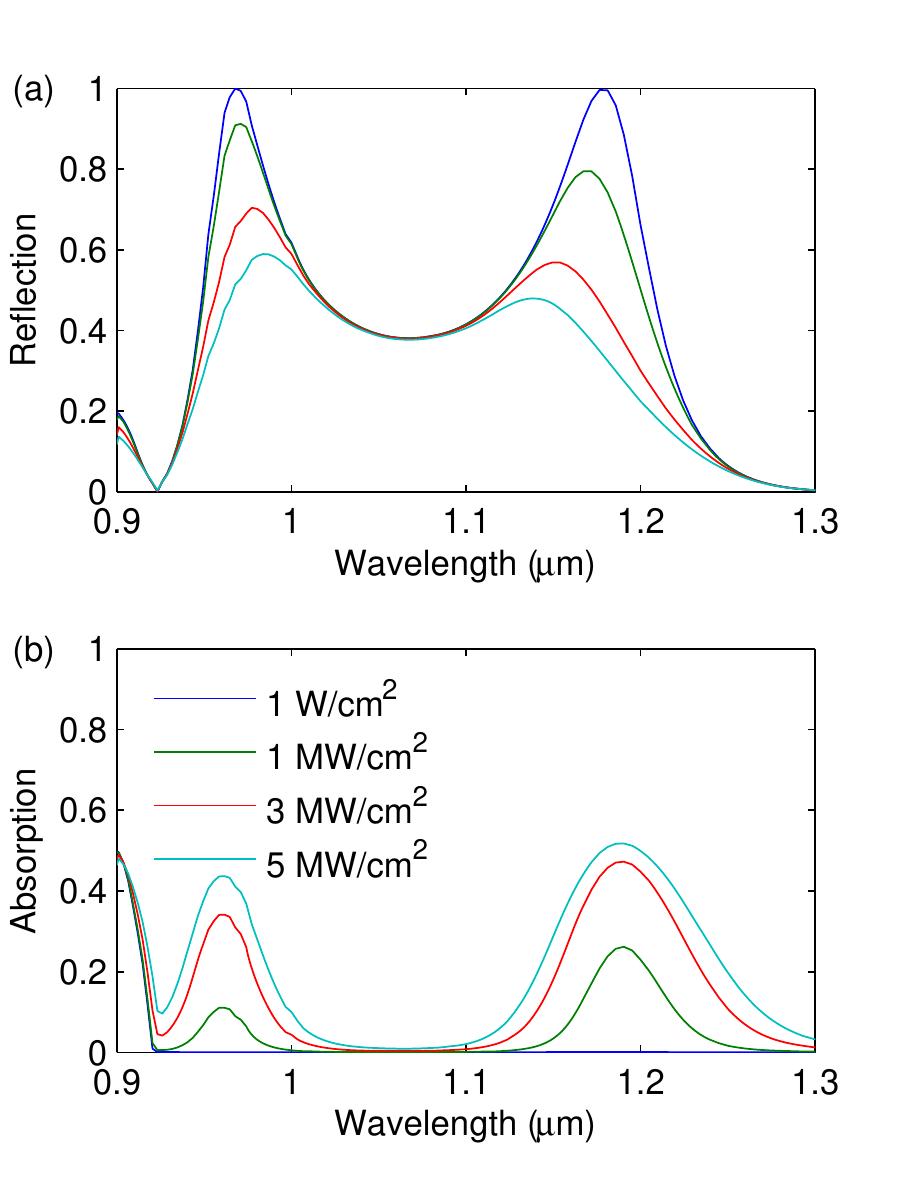}
\caption{\label{fig:epsart} Wavelength dependence of the reflection (a) and absorption (b) for a 0.8 $\mu$m square array of 360-nm diameter InP spheres for different incident intensities.}
\end{figure}

\begin{figure}
\includegraphics[width=63mm]{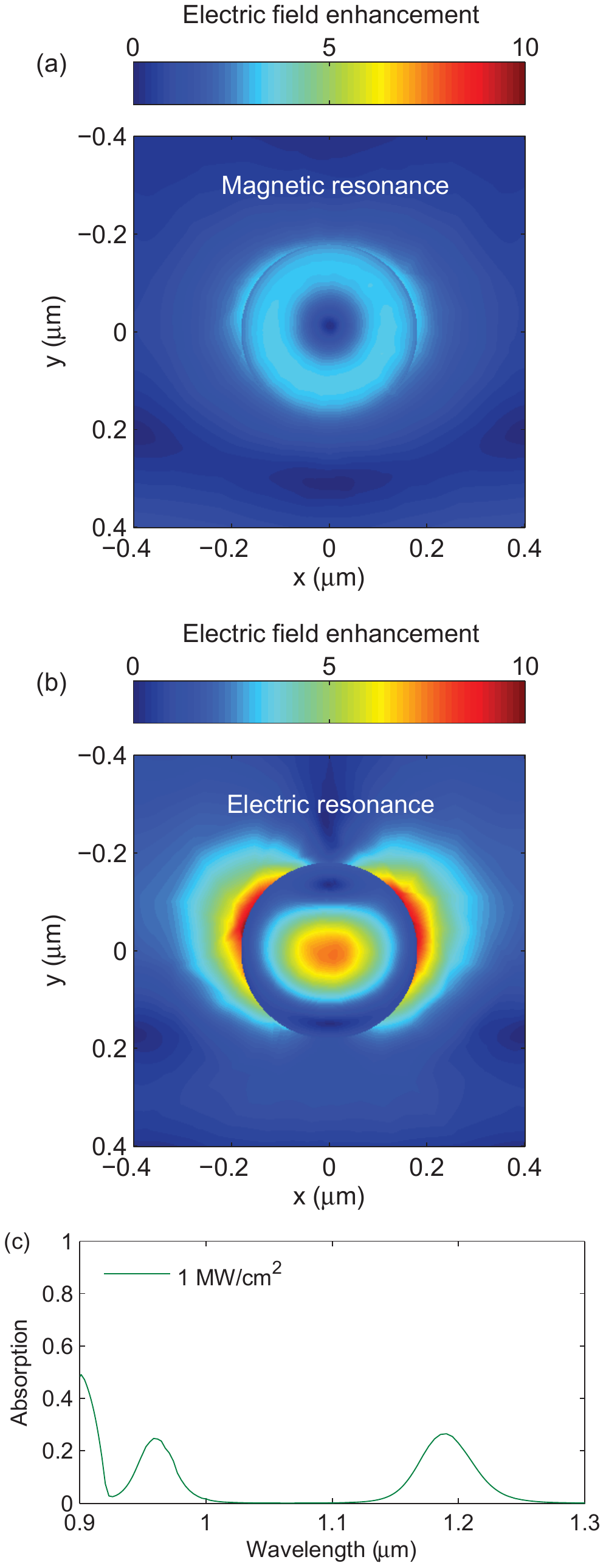}
\caption{\label{fig:epsart} Electric field distributions, normalized to the incident field, at the magnetic resonance (a) and electric resonance (b) for an incident intensity of 1 MW/cm$^2$. (c) Absorption of the metamaterial for an intensity of 1 MW/cm$^2$, assuming a constant free carrier absorption cross section.}
\end{figure}

We note that for a given intensity, the absorption is larger at the magnetic resonance (1.2 $\mu$m) than at the electric resonance (0.95 $\mu$m). This is a surprising result considering the electric field distributions at the resonances, shown in Fig. 6(a) and (b) normalized to the incident electric field for an intensity of 1 MW/cm$^2$. At the magnetic dipole resonance, the electric field is ring shaped and relatively uniform, while the electric field at the electric dipole resonance is highly concentrated at the center of the sphere. Note the field concentration outside the sphere arises from the boundary condition on the normal component electric field, which is discontinuous by the ratio of the dielectric constants of the sphere and free space \cite{Slovick2017}. Thus, based on the field distributions, one might expect the absorption at the electric resonance to be larger due to the larger field concentration. However, we find more absorption at the magnetic resonance. We attribute this to the FCA cross section being about 5 times larger at the magnetic resonance than at the electric resonance [Fig. 3(b)]. To validate this claim, we recalculated the spectral absorption for 1 MW/cm$^2$ intensity, assuming a constant FCA cross section. The results, shown in Fig. 6(c), show that the absorption is approximately equal at the two resonances, confirming that the wavelength-dependent FCA cross section is responsible for the larger absorption at the magnetic resonance. The fact that the absorption is equal at the two resonances for constant FCA is also counterintuitive, since more absorption is expected at the electric resonance because of the larger field concentration.

Since the electric field is enhanced at both resonances, we expect more absorption per unit length in the metamaterial than in a homogeneous material. To illustrate this, we calculated the reflection and absorption of a 360 nm-thick slab of InP, equal in thickness to the InP sphere metamaterial in Fig. 1(a). We see from Fig. 7 that 100 MW/cm$^2$ of intensity is needed to obtain 40\% absorption near 1.2 $\mu$m in the homogenous layer, whereas the metamaterial obtains a similar level of absorption for 3 MW/cm$^2$. Thus, the homogenous layer requires much higher intensities to achieve absorption values comparable to the metamaterial. This factor of 30 higher intensity is consistent with the five-fold field enhancement at the magnetic resonance shown in Fig. 6(a).

\begin{figure}
\includegraphics[width=64mm]{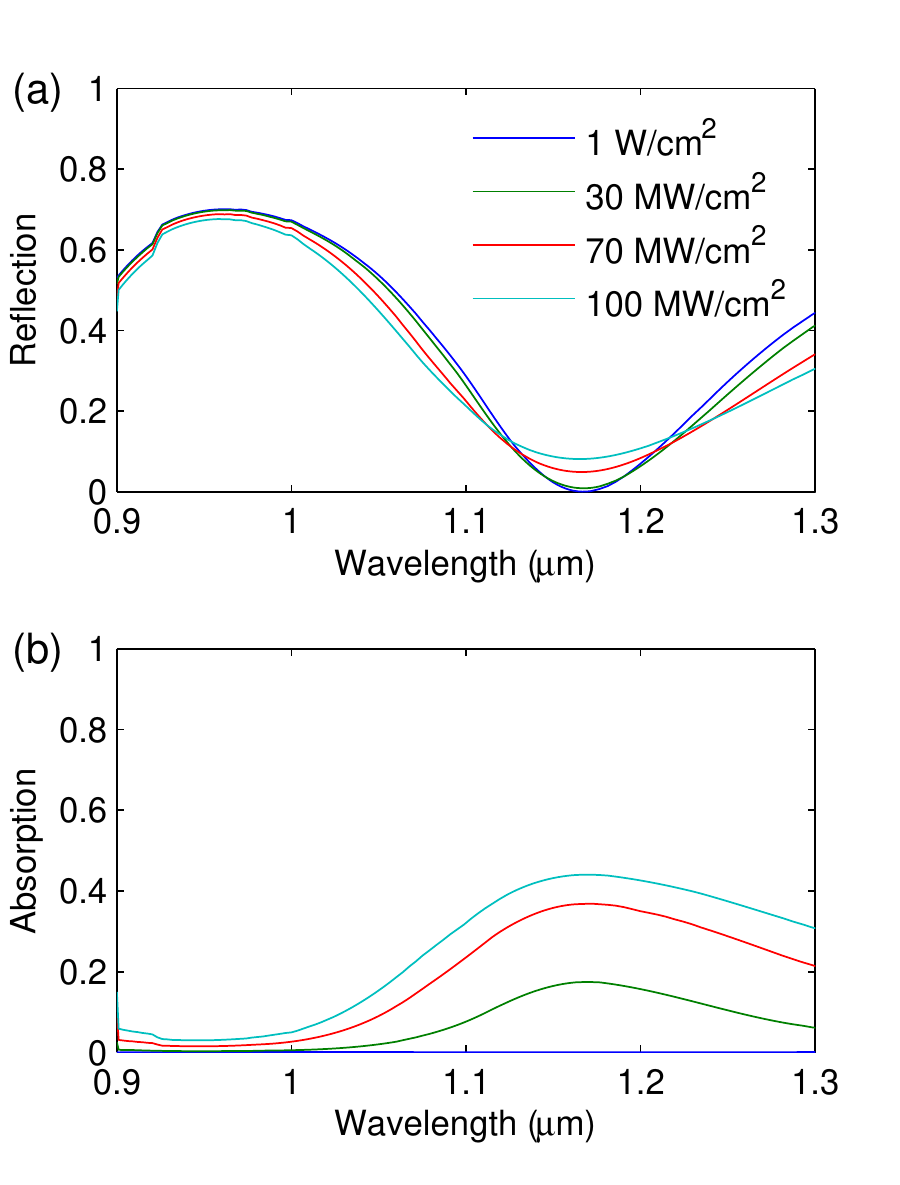}
\caption{\label{fig:epsart} Wavelength dependence of the reflection (a) and absorption (b) for a 360 nm thick film of InP for different incident intensities.}
\end{figure}

\begin{figure}
\includegraphics[width=63mm]{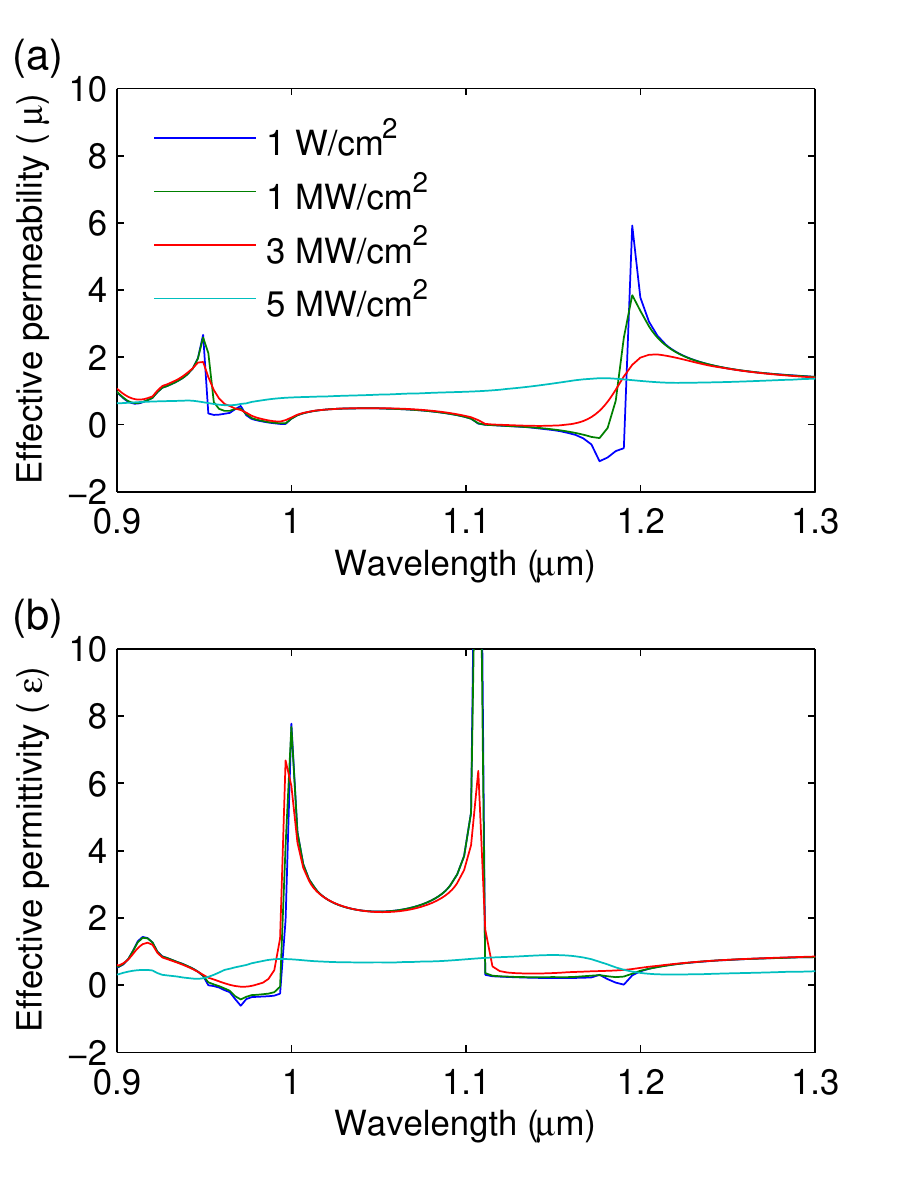}
\caption{\label{fig:epsart} Wavelength dependence of the effective permeability (a) and permittivity (b) of the InP metamaterial for different incident intensities.}
\end{figure}

It is also important to understand the impact of nonlinear absorption on the effective permittivity ($\epsilon$) and permeability ($\mu$) of the metamaterial, as $\epsilon$ and $\mu$ are often used to obtain unique properties such as negative refraction \cite{Smith2000} and perfect reflection \cite{Slovick2013}. The calculated real parts of $\epsilon$ and $\mu$, shown in Fig. 8, were obtained using $S$-parameter inversion, assuming a layer thickness of 1.24 $\mu$m. At low intensities, shown as the blue line in Fig. 8(a), we see a strong resonance in $\mu$ near 1.2 $\mu$m, which arises from the magnetic resonance. The weaker resonance near 0.95 $\mu$m is the anti-resonance associated with the strong electric resonance at that wavelength, clearly seen in Fig. 8(b) for the permittivity. We also find a strong anti-resonance in $\epsilon$ at 1.1 $\mu$m, which arises from the magnetic resonance at 1.2 $\mu$m. These anti-resonances are an artifact of $S$-parameter retrieval that arises from approximating a Bloch wave by a plane wave \cite{Koschny2003}. In the resonance regions, either $\epsilon$ or $\mu$ is negative, resulting in single-negative regions and high reflectivity, as shown in Fig. 5(a). As the intensity increases to 1 MW/cm$^2$ and 3 MW/cm$^2$, we find that the magnetic resonance, near 1.2 $\mu$m in Fig. 8(a), begins to dampen while the electric resonance, near 1 $\mu$m in Fig. 8(b), is largely unchanged. Only when the intensity exceeds 3 MW/cm$^2$ does the electric resonance begin to dampen. For 5 MW/cm$^2$, both resonances are completely dampened. The dampening of the resonances with increasing intensity is also consistent with the decreasing reflection in Fig. 5(a). As the resonance in the real part of $\epsilon$ and $\mu$ broadens, the corresponding imaginary parts of $\epsilon$ and $\mu$ (not shown) also broaden, as per the Kramer-Kronig relationship, which results in broader-band absorption with increasing intensity, as shown in Fig. 5(b). 

In summary, we developed a full-wave model to study the effects of two-photon absorption and photo-induced free-carrier absorption on the effective parameters and optical properties of a structured dielectric metamaterial operating in the near infrared spectral band. As expected, we find that nonlinear absorption leads to dampening of the electric and magnetic Mie resonances at high intensities, with an onset around 1 MW/cm$^2$ for continuous wave illumination. The resonances are almost completely dampened for intensities around 5 MW/cm$^2$. Surprisingly, we find that the nonlinear absorption at the magnetic resonance is larger than at the electric resonance, despite the electric field being more concentrated at the electric resonance. We find this is because the free-carrier absorption cross section is considerably larger at the longer wavelengths near the magnetic resonance. We also find that the metamaterial provides absorption comparable to a homogeneous layer of the same thickness at approximately 30 times less intensity. The lower threshold intensity and smaller footprint for nonlinear absorption can be exploited in applications involving optical limiting, frequency conversion, the Kerr effect, and four-wave mixing.

\section{Acknowledgement} This work was funded by the Office of Naval Research (ONR) through Contract No. N00014-16-C-1023

\bibliography{bib}

\providecommand{\noopsort}[1]{}\providecommand{\singleletter}[1]{#1}%
\begin{thebibliography}{44}%
\makeatletter
\providecommand \@ifxundefined [1]{%
 \@ifx{#1\undefined}
}%
\providecommand \@ifnum [1]{%
 \ifnum #1\expandafter \@firstoftwo
 \else \expandafter \@secondoftwo
 \fi
}%
\providecommand \@ifx [1]{%
 \ifx #1\expandafter \@firstoftwo
 \else \expandafter \@secondoftwo
 \fi
}%
\providecommand \natexlab [1]{#1}%
\providecommand \enquote  [1]{``#1''}%
\providecommand \bibnamefont  [1]{#1}%
\providecommand \bibfnamefont [1]{#1}%
\providecommand \citenamefont [1]{#1}%
\providecommand \href@noop [0]{\@secondoftwo}%
\providecommand \href [0]{\begingroup \@sanitize@url \@href}%
\providecommand \@href[1]{\@@startlink{#1}\@@href}%
\providecommand \@@href[1]{\endgroup#1\@@endlink}%
\providecommand \@sanitize@url [0]{\catcode `\\12\catcode `\$12\catcode
  `\&12\catcode `\#12\catcode `\^12\catcode `\_12\catcode `\%12\relax}%
\providecommand \@@startlink[1]{}%
\providecommand \@@endlink[0]{}%
\providecommand \url  [0]{\begingroup\@sanitize@url \@url }%
\providecommand \@url [1]{\endgroup\@href {#1}{\urlprefix }}%
\providecommand \urlprefix  [0]{URL }%
\providecommand \Eprint [0]{\href }%
\providecommand \doibase [0]{http://dx.doi.org/}%
\providecommand \selectlanguage [0]{\@gobble}%
\providecommand \bibinfo  [0]{\@secondoftwo}%
\providecommand \bibfield  [0]{\@secondoftwo}%
\providecommand \translation [1]{[#1]}%
\providecommand \BibitemOpen [0]{}%
\providecommand \bibitemStop [0]{}%
\providecommand \bibitemNoStop [0]{.\EOS\space}%
\providecommand \EOS [0]{\spacefactor3000\relax}%
\providecommand \BibitemShut  [1]{\csname bibitem#1\endcsname}%
\let\auto@bib@innerbib\@empty
\bibitem [{\citenamefont {Stryland}\ \emph {et~al.}(1985)\citenamefont
  {Stryland}, \citenamefont {Vanherzeele}, \citenamefont {Woodall},
  \citenamefont {Soileau}, \citenamefont {Smirl}, \citenamefont {Guha},\ and\
  \citenamefont {Boggess}}]{Stryland1985}%
  \BibitemOpen
  \bibfield  {author} {\bibinfo {author} {\bibfnamefont {E.~W.~V.}\
  \bibnamefont {Stryland}}, \bibinfo {author} {\bibfnamefont {H.}~\bibnamefont
  {Vanherzeele}}, \bibinfo {author} {\bibfnamefont {M.~A.}\ \bibnamefont
  {Woodall}}, \bibinfo {author} {\bibfnamefont {M.~J.}\ \bibnamefont
  {Soileau}}, \bibinfo {author} {\bibfnamefont {A.~L.}\ \bibnamefont {Smirl}},
  \bibinfo {author} {\bibfnamefont {S.}~\bibnamefont {Guha}}, \ and\ \bibinfo
  {author} {\bibfnamefont {T.~F.}\ \bibnamefont {Boggess}},\ }\href@noop {}
  {\bibfield  {journal} {\bibinfo  {journal} {Opt. Eng.}\ }\textbf {\bibinfo
  {volume} {24}},\ \bibinfo {pages} {244613} (\bibinfo {year}
  {1985})}\BibitemShut {NoStop}%
\bibitem [{\citenamefont {Boggess}\ \emph {et~al.}(1985)\citenamefont
  {Boggess}, \citenamefont {Smirl}, \citenamefont {Moss}, \citenamefont
  {Boyd},\ and\ \citenamefont {Stryland}}]{Boggess1985}%
  \BibitemOpen
  \bibfield  {author} {\bibinfo {author} {\bibfnamefont {T.}~\bibnamefont
  {Boggess}}, \bibinfo {author} {\bibfnamefont {A.}~\bibnamefont {Smirl}},
  \bibinfo {author} {\bibfnamefont {S.}~\bibnamefont {Moss}}, \bibinfo {author}
  {\bibfnamefont {I.}~\bibnamefont {Boyd}}, \ and\ \bibinfo {author}
  {\bibfnamefont {E.~V.}\ \bibnamefont {Stryland}},\ }\href@noop {} {\bibfield
  {journal} {\bibinfo  {journal} {‎IEEE J. Quant. Electron.}\ }\textbf
  {\bibinfo {volume} {21}},\ \bibinfo {pages} {488} (\bibinfo {year}
  {1985})}\BibitemShut {NoStop}%
\bibitem [{\citenamefont {Stryland}\ \emph {et~al.}(1988)\citenamefont
  {Stryland}, \citenamefont {Wu}, \citenamefont {Hagan}, \citenamefont
  {Soileau},\ and\ \citenamefont {Mansour}}]{Stryland1988}%
  \BibitemOpen
  \bibfield  {author} {\bibinfo {author} {\bibfnamefont {E.~W.~V.}\
  \bibnamefont {Stryland}}, \bibinfo {author} {\bibfnamefont {Y.~Y.}\
  \bibnamefont {Wu}}, \bibinfo {author} {\bibfnamefont {D.~J.}\ \bibnamefont
  {Hagan}}, \bibinfo {author} {\bibfnamefont {M.~J.}\ \bibnamefont {Soileau}},
  \ and\ \bibinfo {author} {\bibfnamefont {K.}~\bibnamefont {Mansour}},\
  }\href@noop {} {\bibfield  {journal} {\bibinfo  {journal} {‎JOSA B}\
  }\textbf {\bibinfo {volume} {5}},\ \bibinfo {pages} {1980} (\bibinfo {year}
  {1988})}\BibitemShut {NoStop}%
\bibitem [{\citenamefont {Scalora}\ \emph {et~al.}(1994)\citenamefont
  {Scalora}, \citenamefont {Dowling}, \citenamefont {Bowden},\ and\
  \citenamefont {Bloemer}}]{Scalora1994}%
  \BibitemOpen
  \bibfield  {author} {\bibinfo {author} {\bibfnamefont {M.}~\bibnamefont
  {Scalora}}, \bibinfo {author} {\bibfnamefont {J.~P.}\ \bibnamefont
  {Dowling}}, \bibinfo {author} {\bibfnamefont {C.~M.}\ \bibnamefont {Bowden}},
  \ and\ \bibinfo {author} {\bibfnamefont {M.~J.}\ \bibnamefont {Bloemer}},\
  }\href@noop {} {\bibfield  {journal} {\bibinfo  {journal} {‎Phys.Rev.
  Lett.}\ }\textbf {\bibinfo {volume} {73}},\ \bibinfo {pages} {1368} (\bibinfo
  {year} {1994})}\BibitemShut {NoStop}%
\bibitem [{\citenamefont {Tran}(1997)}]{Tran1997}%
  \BibitemOpen
  \bibfield  {author} {\bibinfo {author} {\bibfnamefont {P.}~\bibnamefont
  {Tran}},\ }\href@noop {} {\bibfield  {journal} {\bibinfo  {journal} {‎JOSA
  B}\ }\textbf {\bibinfo {volume} {14}},\ \bibinfo {pages} {2589} (\bibinfo
  {year} {1997})}\BibitemShut {NoStop}%
\bibitem [{\citenamefont {Hache}\ and\ \citenamefont
  {Bourgeois}(2000)}]{Hache2000}%
  \BibitemOpen
  \bibfield  {author} {\bibinfo {author} {\bibfnamefont {A.}~\bibnamefont
  {Hache}}\ and\ \bibinfo {author} {\bibfnamefont {M.}~\bibnamefont
  {Bourgeois}},\ }\href@noop {} {\bibfield  {journal} {\bibinfo  {journal}
  {‎Appl. Phys. Lett.}\ }\textbf {\bibinfo {volume} {77}},\ \bibinfo {pages}
  {4089} (\bibinfo {year} {2000})}\BibitemShut {NoStop}%
\bibitem [{\citenamefont {Rashkeev}\ and\ \citenamefont
  {Lambrecht}(2001)}]{Rashkeev2001}%
  \BibitemOpen
  \bibfield  {author} {\bibinfo {author} {\bibfnamefont {S.~N.}\ \bibnamefont
  {Rashkeev}}\ and\ \bibinfo {author} {\bibfnamefont {W.~R.}\ \bibnamefont
  {Lambrecht}},\ }\href@noop {} {\bibfield  {journal} {\bibinfo  {journal}
  {‎Phys. Rev. B}\ }\textbf {\bibinfo {volume} {63}},\ \bibinfo {pages}
  {165212} (\bibinfo {year} {2001})}\BibitemShut {NoStop}%
\bibitem [{\citenamefont {Lekse}\ \emph {et~al.}(2009)\citenamefont {Lekse},
  \citenamefont {Moreau}, \citenamefont {McNerny}, \citenamefont {Yeon},
  \citenamefont {Halasyamani},\ and\ \citenamefont {Aitken}}]{Lekse2009}%
  \BibitemOpen
  \bibfield  {author} {\bibinfo {author} {\bibfnamefont {J.~M.}\ \bibnamefont
  {Lekse}}, \bibinfo {author} {\bibfnamefont {M.~A.}\ \bibnamefont {Moreau}},
  \bibinfo {author} {\bibfnamefont {K.~L.}\ \bibnamefont {McNerny}}, \bibinfo
  {author} {\bibfnamefont {J.}~\bibnamefont {Yeon}}, \bibinfo {author}
  {\bibfnamefont {P.~S.}\ \bibnamefont {Halasyamani}}, \ and\ \bibinfo {author}
  {\bibfnamefont {J.~A.}\ \bibnamefont {Aitken}},\ }\href@noop {} {\bibfield
  {journal} {\bibinfo  {journal} {‎Inorg. Chem.}\ }\textbf {\bibinfo {volume}
  {48}},\ \bibinfo {pages} {7516} (\bibinfo {year} {2009})}\BibitemShut
  {NoStop}%
\bibitem [{\citenamefont {Gallo}\ and\ \citenamefont
  {Assanto}(1999)}]{Gallo1999}%
  \BibitemOpen
  \bibfield  {author} {\bibinfo {author} {\bibfnamefont {K.}~\bibnamefont
  {Gallo}}\ and\ \bibinfo {author} {\bibfnamefont {G.}~\bibnamefont
  {Assanto}},\ }\href@noop {} {\bibfield  {journal} {\bibinfo  {journal}
  {‎JOSA B}\ }\textbf {\bibinfo {volume} {16}},\ \bibinfo {pages} {267}
  (\bibinfo {year} {1999})}\BibitemShut {NoStop}%
\bibitem [{\citenamefont {Yu}\ and\ \citenamefont {Fan}(2009)}]{Yu2009}%
  \BibitemOpen
  \bibfield  {author} {\bibinfo {author} {\bibfnamefont {Z.}~\bibnamefont
  {Yu}}\ and\ \bibinfo {author} {\bibfnamefont {S.}~\bibnamefont {Fan}},\
  }\href@noop {} {\bibfield  {journal} {\bibinfo  {journal} {‎Nat. Photon.}\
  }\textbf {\bibinfo {volume} {3}},\ \bibinfo {pages} {91} (\bibinfo {year}
  {2009})}\BibitemShut {NoStop}%
\bibitem [{\citenamefont {Chang}\ \emph {et~al.}(2014)\citenamefont {Chang},
  \citenamefont {Vuletic},\ and\ \citenamefont {Lukin}}]{Chang2014}%
  \BibitemOpen
  \bibfield  {author} {\bibinfo {author} {\bibfnamefont {D.~E.}\ \bibnamefont
  {Chang}}, \bibinfo {author} {\bibfnamefont {V.}~\bibnamefont {Vuletic}}, \
  and\ \bibinfo {author} {\bibfnamefont {M.~D.}\ \bibnamefont {Lukin}},\
  }\href@noop {} {\bibfield  {journal} {\bibinfo  {journal} {‎Nat. Photon.}\
  }\textbf {\bibinfo {volume} {8}},\ \bibinfo {pages} {685} (\bibinfo {year}
  {2014})}\BibitemShut {NoStop}%
\bibitem [{\citenamefont {Bechtel}\ and\ \citenamefont
  {Smith}(1976)}]{Bechtel1976}%
  \BibitemOpen
  \bibfield  {author} {\bibinfo {author} {\bibfnamefont {J.~H.}\ \bibnamefont
  {Bechtel}}\ and\ \bibinfo {author} {\bibfnamefont {W.~L.}\ \bibnamefont
  {Smith}},\ }\href@noop {} {\bibfield  {journal} {\bibinfo  {journal}
  {‎Phys. Rev. B}\ }\textbf {\bibinfo {volume} {13}},\ \bibinfo {pages}
  {3515} (\bibinfo {year} {1976})}\BibitemShut {NoStop}%
\bibitem [{\citenamefont {Boggess}\ \emph {et~al.}(1986)\citenamefont
  {Boggess}, \citenamefont {Bohnert}, \citenamefont {Mansour}, \citenamefont
  {Moss}, \citenamefont {Boyd},\ and\ \citenamefont {Smirl}}]{Boggess1986}%
  \BibitemOpen
  \bibfield  {author} {\bibinfo {author} {\bibfnamefont {T.}~\bibnamefont
  {Boggess}}, \bibinfo {author} {\bibfnamefont {K.}~\bibnamefont {Bohnert}},
  \bibinfo {author} {\bibfnamefont {K.}~\bibnamefont {Mansour}}, \bibinfo
  {author} {\bibfnamefont {S.}~\bibnamefont {Moss}}, \bibinfo {author}
  {\bibfnamefont {I.}~\bibnamefont {Boyd}}, \ and\ \bibinfo {author}
  {\bibfnamefont {A.}~\bibnamefont {Smirl}},\ }\href@noop {} {\bibfield
  {journal} {\bibinfo  {journal} {IEEE J. Quant. Electron.}\ }\textbf {\bibinfo
  {volume} {22}},\ \bibinfo {pages} {360} (\bibinfo {year} {1986})}\BibitemShut
  {NoStop}%
\bibitem [{\citenamefont {Krishnamurthy}\ \emph {et~al.}(2011)\citenamefont
  {Krishnamurthy}, \citenamefont {Yu}, \citenamefont {Gonzalez},\ and\
  \citenamefont {Guha}}]{Krishnamurthy2011}%
  \BibitemOpen
  \bibfield  {author} {\bibinfo {author} {\bibfnamefont {S.}~\bibnamefont
  {Krishnamurthy}}, \bibinfo {author} {\bibfnamefont {Z.~G.}\ \bibnamefont
  {Yu}}, \bibinfo {author} {\bibfnamefont {L.~P.}\ \bibnamefont {Gonzalez}}, \
  and\ \bibinfo {author} {\bibfnamefont {S.}~\bibnamefont {Guha}},\ }\href@noop
  {} {\bibfield  {journal} {\bibinfo  {journal} {J. Appl. Phys.}\ }\textbf
  {\bibinfo {volume} {109}},\ \bibinfo {pages} {033102} (\bibinfo {year}
  {2011})}\BibitemShut {NoStop}%
\bibitem [{\citenamefont {Kauranen}\ and\ \citenamefont
  {Zayats}(2012)}]{Kauranen2012}%
  \BibitemOpen
  \bibfield  {author} {\bibinfo {author} {\bibfnamefont {M.}~\bibnamefont
  {Kauranen}}\ and\ \bibinfo {author} {\bibfnamefont {A.~V.}\ \bibnamefont
  {Zayats}},\ }\href@noop {} {\bibfield  {journal} {\bibinfo  {journal} {Nat.
  Photon.}\ }\textbf {\bibinfo {volume} {737}},\ \bibinfo {pages} {6} (\bibinfo
  {year} {2012})}\BibitemShut {NoStop}%
\bibitem [{\citenamefont {Schuller}\ \emph {et~al.}(2010)\citenamefont
  {Schuller}, \citenamefont {Barnard}, \citenamefont {Cai}, \citenamefont
  {Jun}, \citenamefont {White},\ and\ \citenamefont
  {Brongersma}}]{Schuller2010}%
  \BibitemOpen
  \bibfield  {author} {\bibinfo {author} {\bibfnamefont {J.~A.}\ \bibnamefont
  {Schuller}}, \bibinfo {author} {\bibfnamefont {E.~S.}\ \bibnamefont
  {Barnard}}, \bibinfo {author} {\bibfnamefont {W.}~\bibnamefont {Cai}},
  \bibinfo {author} {\bibfnamefont {Y.~C.}\ \bibnamefont {Jun}}, \bibinfo
  {author} {\bibfnamefont {J.~S.}\ \bibnamefont {White}}, \ and\ \bibinfo
  {author} {\bibfnamefont {M.~L.}\ \bibnamefont {Brongersma}},\ }\href@noop {}
  {\bibfield  {journal} {\bibinfo  {journal} {Nat. Mater.}\ }\textbf {\bibinfo
  {volume} {9}},\ \bibinfo {pages} {193} (\bibinfo {year} {2010})}\BibitemShut
  {NoStop}%
\bibitem [{\citenamefont {Czaplicki}\ \emph {et~al.}(2013)\citenamefont
  {Czaplicki}, \citenamefont {Husu}, \citenamefont {Siikanen}, \citenamefont
  {Makitalo}, \citenamefont {Kauranen}, \citenamefont {Laukkanen},
  \citenamefont {Lehtolahti},\ and\ \citenamefont {Kuittinen}}]{Czaplicki2013}%
  \BibitemOpen
  \bibfield  {author} {\bibinfo {author} {\bibfnamefont {R.}~\bibnamefont
  {Czaplicki}}, \bibinfo {author} {\bibfnamefont {H.}~\bibnamefont {Husu}},
  \bibinfo {author} {\bibfnamefont {R.}~\bibnamefont {Siikanen}}, \bibinfo
  {author} {\bibfnamefont {J.}~\bibnamefont {Makitalo}}, \bibinfo {author}
  {\bibfnamefont {M.}~\bibnamefont {Kauranen}}, \bibinfo {author}
  {\bibfnamefont {J.}~\bibnamefont {Laukkanen}}, \bibinfo {author}
  {\bibfnamefont {J.}~\bibnamefont {Lehtolahti}}, \ and\ \bibinfo {author}
  {\bibfnamefont {M.}~\bibnamefont {Kuittinen}},\ }\href@noop {} {\bibfield
  {journal} {\bibinfo  {journal} {Phys. Rev. Lett.}\ }\textbf {\bibinfo
  {volume} {110}},\ \bibinfo {pages} {093902} (\bibinfo {year}
  {2013})}\BibitemShut {NoStop}%
\bibitem [{\citenamefont {Aouani}\ \emph {et~al.}(2012)\citenamefont {Aouani},
  \citenamefont {Navarro-Cia}, \citenamefont {Rahmani}, \citenamefont
  {Sidiropoulos}, \citenamefont {Hong}, \citenamefont {Oulton},\ and\
  \citenamefont {Maier}}]{Aouani2012}%
  \BibitemOpen
  \bibfield  {author} {\bibinfo {author} {\bibfnamefont {H.}~\bibnamefont
  {Aouani}}, \bibinfo {author} {\bibfnamefont {M.}~\bibnamefont {Navarro-Cia}},
  \bibinfo {author} {\bibfnamefont {M.}~\bibnamefont {Rahmani}}, \bibinfo
  {author} {\bibfnamefont {T.~P.}\ \bibnamefont {Sidiropoulos}}, \bibinfo
  {author} {\bibfnamefont {M.}~\bibnamefont {Hong}}, \bibinfo {author}
  {\bibfnamefont {R.~F.}\ \bibnamefont {Oulton}}, \ and\ \bibinfo {author}
  {\bibfnamefont {S.~A.}\ \bibnamefont {Maier}},\ }\href@noop {} {\bibfield
  {journal} {\bibinfo  {journal} {Nano Lett.}\ }\textbf {\bibinfo {volume}
  {12}},\ \bibinfo {pages} {4997} (\bibinfo {year} {2012})}\BibitemShut
  {NoStop}%
\bibitem [{\citenamefont {Thyagarajan}\ \emph {et~al.}(2012)\citenamefont
  {Thyagarajan}, \citenamefont {Rivier}, \citenamefont {Lovera},\ and\
  \citenamefont {Martin}}]{Thyagarajan2012}%
  \BibitemOpen
  \bibfield  {author} {\bibinfo {author} {\bibfnamefont {K.}~\bibnamefont
  {Thyagarajan}}, \bibinfo {author} {\bibfnamefont {S.}~\bibnamefont {Rivier}},
  \bibinfo {author} {\bibfnamefont {A.}~\bibnamefont {Lovera}}, \ and\ \bibinfo
  {author} {\bibfnamefont {O.~J.}\ \bibnamefont {Martin}},\ }\href@noop {}
  {\bibfield  {journal} {\bibinfo  {journal} {Opt. Exp.}\ }\textbf {\bibinfo
  {volume} {20}},\ \bibinfo {pages} {12860} (\bibinfo {year}
  {2012})}\BibitemShut {NoStop}%
\bibitem [{\citenamefont {Thyagarajan}\ \emph {et~al.}(2013)\citenamefont
  {Thyagarajan}, \citenamefont {Butet},\ and\ \citenamefont
  {Martin}}]{Thyagarajan2013}%
  \BibitemOpen
  \bibfield  {author} {\bibinfo {author} {\bibfnamefont {K.}~\bibnamefont
  {Thyagarajan}}, \bibinfo {author} {\bibfnamefont {J.}~\bibnamefont {Butet}},
  \ and\ \bibinfo {author} {\bibfnamefont {O.~J.}\ \bibnamefont {Martin}},\
  }\href@noop {} {\bibfield  {journal} {\bibinfo  {journal} {Nano Lett.}\
  }\textbf {\bibinfo {volume} {13}},\ \bibinfo {pages} {1847} (\bibinfo {year}
  {2013})}\BibitemShut {NoStop}%
\bibitem [{\citenamefont {Zhang}\ \emph {et~al.}(2011)\citenamefont {Zhang},
  \citenamefont {Grady}, \citenamefont {Ayala-Orozco},\ and\ \citenamefont
  {Halas}}]{Zhang2011}%
  \BibitemOpen
  \bibfield  {author} {\bibinfo {author} {\bibfnamefont {Y.}~\bibnamefont
  {Zhang}}, \bibinfo {author} {\bibfnamefont {N.~K.}\ \bibnamefont {Grady}},
  \bibinfo {author} {\bibfnamefont {C.}~\bibnamefont {Ayala-Orozco}}, \ and\
  \bibinfo {author} {\bibfnamefont {N.~J.}\ \bibnamefont {Halas}},\ }\href@noop
  {} {\bibfield  {journal} {\bibinfo  {journal} {Nano Lett.}\ }\textbf
  {\bibinfo {volume} {11}},\ \bibinfo {pages} {5519} (\bibinfo {year}
  {2011})}\BibitemShut {NoStop}%
\bibitem [{\citenamefont {Harutyunyan}\ \emph {et~al.}(2012)\citenamefont
  {Harutyunyan}, \citenamefont {Volpe}, \citenamefont {Quidant},\ and\
  \citenamefont {Novotny}}]{Harutyunyan2012}%
  \BibitemOpen
  \bibfield  {author} {\bibinfo {author} {\bibfnamefont {H.}~\bibnamefont
  {Harutyunyan}}, \bibinfo {author} {\bibfnamefont {G.}~\bibnamefont {Volpe}},
  \bibinfo {author} {\bibfnamefont {R.}~\bibnamefont {Quidant}}, \ and\
  \bibinfo {author} {\bibfnamefont {L.}~\bibnamefont {Novotny}},\ }\href@noop
  {} {\bibfield  {journal} {\bibinfo  {journal} {Phys. Rev. Lett.}\ }\textbf
  {\bibinfo {volume} {108}},\ \bibinfo {pages} {217403} (\bibinfo {year}
  {2012})}\BibitemShut {NoStop}%
\bibitem [{\citenamefont {Navarro-Cia}\ and\ \citenamefont
  {Maier}(2012)}]{Navarro2012}%
  \BibitemOpen
  \bibfield  {author} {\bibinfo {author} {\bibfnamefont {M.}~\bibnamefont
  {Navarro-Cia}}\ and\ \bibinfo {author} {\bibfnamefont {S.~A.}\ \bibnamefont
  {Maier}},\ }\href@noop {} {\bibfield  {journal} {\bibinfo  {journal} {ACS
  Nano}\ }\textbf {\bibinfo {volume} {6}},\ \bibinfo {pages} {3537} (\bibinfo
  {year} {2012})}\BibitemShut {NoStop}%
\bibitem [{\citenamefont {Gu}\ \emph {et~al.}(2012)\citenamefont {Gu},
  \citenamefont {Singh}, \citenamefont {Liu}, \citenamefont {Zhang},
  \citenamefont {Ma}, \citenamefont {Zhang}, \citenamefont {Maier},
  \citenamefont {Tian}, \citenamefont {Azad}, \citenamefont {Chen},\ and\
  \citenamefont {Taylor}}]{Gu2012}%
  \BibitemOpen
  \bibfield  {author} {\bibinfo {author} {\bibfnamefont {J.}~\bibnamefont
  {Gu}}, \bibinfo {author} {\bibfnamefont {R.}~\bibnamefont {Singh}}, \bibinfo
  {author} {\bibfnamefont {X.}~\bibnamefont {Liu}}, \bibinfo {author}
  {\bibfnamefont {X.}~\bibnamefont {Zhang}}, \bibinfo {author} {\bibfnamefont
  {Y.}~\bibnamefont {Ma}}, \bibinfo {author} {\bibfnamefont {S.}~\bibnamefont
  {Zhang}}, \bibinfo {author} {\bibfnamefont {S.~A.}\ \bibnamefont {Maier}},
  \bibinfo {author} {\bibfnamefont {Z.}~\bibnamefont {Tian}}, \bibinfo {author}
  {\bibfnamefont {A.~K.}\ \bibnamefont {Azad}}, \bibinfo {author}
  {\bibfnamefont {H.~T.}\ \bibnamefont {Chen}}, \ and\ \bibinfo {author}
  {\bibfnamefont {A.~J.}\ \bibnamefont {Taylor}},\ }\href@noop {} {\bibfield
  {journal} {\bibinfo  {journal} {Nat. Comm.}\ }\textbf {\bibinfo {volume}
  {3}},\ \bibinfo {pages} {1151} (\bibinfo {year} {2012})}\BibitemShut
  {NoStop}%
\bibitem [{\citenamefont {Kurter}\ \emph {et~al.}(2011)\citenamefont {Kurter},
  \citenamefont {Tassin}, \citenamefont {Zhang}, \citenamefont {Koschny},
  \citenamefont {Zhuravel}, \citenamefont {Ustinov}, \citenamefont {Anlage},\
  and\ \citenamefont {Soukoulis}}]{Kurter2011}%
  \BibitemOpen
  \bibfield  {author} {\bibinfo {author} {\bibfnamefont {C.}~\bibnamefont
  {Kurter}}, \bibinfo {author} {\bibfnamefont {P.}~\bibnamefont {Tassin}},
  \bibinfo {author} {\bibfnamefont {L.}~\bibnamefont {Zhang}}, \bibinfo
  {author} {\bibfnamefont {T.}~\bibnamefont {Koschny}}, \bibinfo {author}
  {\bibfnamefont {A.~P.}\ \bibnamefont {Zhuravel}}, \bibinfo {author}
  {\bibfnamefont {A.~V.}\ \bibnamefont {Ustinov}}, \bibinfo {author}
  {\bibfnamefont {S.~M.}\ \bibnamefont {Anlage}}, \ and\ \bibinfo {author}
  {\bibfnamefont {C.~M.}\ \bibnamefont {Soukoulis}},\ }\href@noop {} {\bibfield
   {journal} {\bibinfo  {journal} {Phys. Rev. Lett.}\ }\textbf {\bibinfo
  {volume} {107}},\ \bibinfo {pages} {043901} (\bibinfo {year}
  {2011})}\BibitemShut {NoStop}%
\bibitem [{\citenamefont {Shcherbakov}\ \emph {et~al.}(2014)\citenamefont
  {Shcherbakov}, \citenamefont {Neshev}, \citenamefont {Hopkins}, \citenamefont
  {Shorokhov}, \citenamefont {Staude}, \citenamefont {Melik-Gaykazyan},
  \citenamefont {Decker}, \citenamefont {Ezhov}, \citenamefont
  {Miroshnichenko}, \citenamefont {Brener},\ and\ \citenamefont
  {Fedyanin}}]{Shcherbakov2014}%
  \BibitemOpen
  \bibfield  {author} {\bibinfo {author} {\bibfnamefont {M.~R.}\ \bibnamefont
  {Shcherbakov}}, \bibinfo {author} {\bibfnamefont {D.~N.}\ \bibnamefont
  {Neshev}}, \bibinfo {author} {\bibfnamefont {B.}~\bibnamefont {Hopkins}},
  \bibinfo {author} {\bibfnamefont {A.~S.}\ \bibnamefont {Shorokhov}}, \bibinfo
  {author} {\bibfnamefont {I.}~\bibnamefont {Staude}}, \bibinfo {author}
  {\bibfnamefont {E.~V.}\ \bibnamefont {Melik-Gaykazyan}}, \bibinfo {author}
  {\bibfnamefont {M.}~\bibnamefont {Decker}}, \bibinfo {author} {\bibfnamefont
  {A.~A.}\ \bibnamefont {Ezhov}}, \bibinfo {author} {\bibfnamefont {A.~E.}\
  \bibnamefont {Miroshnichenko}}, \bibinfo {author} {\bibfnamefont
  {I.}~\bibnamefont {Brener}}, \ and\ \bibinfo {author} {\bibfnamefont {A.~A.}\
  \bibnamefont {Fedyanin}},\ }\href@noop {} {\bibfield  {journal} {\bibinfo
  {journal} {Nano Lett.}\ }\textbf {\bibinfo {volume} {14}},\ \bibinfo {pages}
  {6488} (\bibinfo {year} {2014})}\BibitemShut {NoStop}%
\bibitem [{\citenamefont {Yang}\ \emph {et~al.}(2015)\citenamefont {Yang},
  \citenamefont {Wang}, \citenamefont {Boulesbaa}, \citenamefont {Kravchenko},
  \citenamefont {Briggs}, \citenamefont {Puretzky}, \citenamefont {Geohegan},\
  and\ \citenamefont {Valentine}}]{Yang2015}%
  \BibitemOpen
  \bibfield  {author} {\bibinfo {author} {\bibfnamefont {Y.}~\bibnamefont
  {Yang}}, \bibinfo {author} {\bibfnamefont {W.}~\bibnamefont {Wang}}, \bibinfo
  {author} {\bibfnamefont {A.}~\bibnamefont {Boulesbaa}}, \bibinfo {author}
  {\bibfnamefont {I.~I.}\ \bibnamefont {Kravchenko}}, \bibinfo {author}
  {\bibfnamefont {D.~P.}\ \bibnamefont {Briggs}}, \bibinfo {author}
  {\bibfnamefont {A.}~\bibnamefont {Puretzky}}, \bibinfo {author}
  {\bibfnamefont {D.}~\bibnamefont {Geohegan}}, \ and\ \bibinfo {author}
  {\bibfnamefont {J.}~\bibnamefont {Valentine}},\ }\href@noop {} {\bibfield
  {journal} {\bibinfo  {journal} {Nano Lett.}\ }\textbf {\bibinfo {volume}
  {15}},\ \bibinfo {pages} {7388} (\bibinfo {year} {2015})}\BibitemShut
  {NoStop}%
\bibitem [{\citenamefont {Kapitanova}\ \emph {et~al.}(2017)\citenamefont
  {Kapitanova}, \citenamefont {Ternovski}, \citenamefont {Miroshnichenko},
  \citenamefont {Pavlov}, \citenamefont {Belov}, \citenamefont {Kivshar},\ and\
  \citenamefont {Tribelsky}}]{Kapitanova2017}%
  \BibitemOpen
  \bibfield  {author} {\bibinfo {author} {\bibfnamefont {P.}~\bibnamefont
  {Kapitanova}}, \bibinfo {author} {\bibfnamefont {V.}~\bibnamefont
  {Ternovski}}, \bibinfo {author} {\bibfnamefont {A.}~\bibnamefont
  {Miroshnichenko}}, \bibinfo {author} {\bibfnamefont {N.}~\bibnamefont
  {Pavlov}}, \bibinfo {author} {\bibfnamefont {P.}~\bibnamefont {Belov}},
  \bibinfo {author} {\bibfnamefont {Y.}~\bibnamefont {Kivshar}}, \ and\
  \bibinfo {author} {\bibfnamefont {M.}~\bibnamefont {Tribelsky}},\ }\href@noop
  {} {\bibfield  {journal} {\bibinfo  {journal} {‎Sci. Rep.}\ }\textbf
  {\bibinfo {volume} {7}},\ \bibinfo {pages} {731} (\bibinfo {year}
  {2017})}\BibitemShut {NoStop}%
\bibitem [{\citenamefont {Shcherbakov}\ \emph
  {et~al.}(2015{\natexlab{a}})\citenamefont {Shcherbakov}, \citenamefont
  {Shorokhov}, \citenamefont {Neshev}, \citenamefont {Hopkins}, \citenamefont
  {Staude}, \citenamefont {Melik-Gaykazyan}, \citenamefont {Ezhov},
  \citenamefont {Miroshnichenko}, \citenamefont {Brener}, \citenamefont
  {Fedyanin},\ and\ \citenamefont {Kivshar}}]{Shcherbakov2015a}%
  \BibitemOpen
  \bibfield  {author} {\bibinfo {author} {\bibfnamefont {M.~R.}\ \bibnamefont
  {Shcherbakov}}, \bibinfo {author} {\bibfnamefont {A.~S.}\ \bibnamefont
  {Shorokhov}}, \bibinfo {author} {\bibfnamefont {D.~N.}\ \bibnamefont
  {Neshev}}, \bibinfo {author} {\bibfnamefont {B.}~\bibnamefont {Hopkins}},
  \bibinfo {author} {\bibfnamefont {I.}~\bibnamefont {Staude}}, \bibinfo
  {author} {\bibfnamefont {E.~V.}\ \bibnamefont {Melik-Gaykazyan}}, \bibinfo
  {author} {\bibfnamefont {A.~A.}\ \bibnamefont {Ezhov}}, \bibinfo {author}
  {\bibfnamefont {A.~E.}\ \bibnamefont {Miroshnichenko}}, \bibinfo {author}
  {\bibfnamefont {I.}~\bibnamefont {Brener}}, \bibinfo {author} {\bibfnamefont
  {A.~A.}\ \bibnamefont {Fedyanin}}, \ and\ \bibinfo {author} {\bibfnamefont
  {Y.~S.}\ \bibnamefont {Kivshar}},\ }\href@noop {} {\bibfield  {journal}
  {\bibinfo  {journal} {ACS Photon.}\ }\textbf {\bibinfo {volume} {2}},\
  \bibinfo {pages} {578} (\bibinfo {year} {2015}{\natexlab{a}})}\BibitemShut
  {NoStop}%
\bibitem [{\citenamefont {Smirnova}\ \emph {et~al.}(2016)\citenamefont
  {Smirnova}, \citenamefont {Khanikaev}, \citenamefont {Smirnov},\ and\
  \citenamefont {Kivshar}}]{Smirnova2016}%
  \BibitemOpen
  \bibfield  {author} {\bibinfo {author} {\bibfnamefont {D.~A.}\ \bibnamefont
  {Smirnova}}, \bibinfo {author} {\bibfnamefont {A.~B.}\ \bibnamefont
  {Khanikaev}}, \bibinfo {author} {\bibfnamefont {L.~A.}\ \bibnamefont
  {Smirnov}}, \ and\ \bibinfo {author} {\bibfnamefont {Y.~S.}\ \bibnamefont
  {Kivshar}},\ }\href@noop {} {\bibfield  {journal} {\bibinfo  {journal} {ACS
  Photon.}\ }\textbf {\bibinfo {volume} {3}},\ \bibinfo {pages} {1468}
  (\bibinfo {year} {2016})}\BibitemShut {NoStop}%
\bibitem [{\citenamefont {Grinblat}\ \emph {et~al.}(2016)\citenamefont
  {Grinblat}, \citenamefont {Li}, \citenamefont {Nielsen}, \citenamefont
  {Oulton},\ and\ \citenamefont {Maier}}]{Grinblat2016}%
  \BibitemOpen
  \bibfield  {author} {\bibinfo {author} {\bibfnamefont {G.}~\bibnamefont
  {Grinblat}}, \bibinfo {author} {\bibfnamefont {Y.}~\bibnamefont {Li}},
  \bibinfo {author} {\bibfnamefont {M.~P.}\ \bibnamefont {Nielsen}}, \bibinfo
  {author} {\bibfnamefont {R.~F.}\ \bibnamefont {Oulton}}, \ and\ \bibinfo
  {author} {\bibfnamefont {S.~A.}\ \bibnamefont {Maier}},\ }\href@noop {}
  {\bibfield  {journal} {\bibinfo  {journal} {Nano Lett.}\ }\textbf {\bibinfo
  {volume} {16}},\ \bibinfo {pages} {4635} (\bibinfo {year}
  {2016})}\BibitemShut {NoStop}%
\bibitem [{\citenamefont {Shorokhov}\ \emph {et~al.}(2016)\citenamefont
  {Shorokhov}, \citenamefont {Melik-Gaykazyan}, \citenamefont {Smirnova},
  \citenamefont {Hopkins}, \citenamefont {Chong}, \citenamefont {Choi},
  \citenamefont {Shcherbakov}, \citenamefont {Miroshnichenko}, \citenamefont
  {Neshev}, \citenamefont {Fedyanin},\ and\ \citenamefont
  {Kivshar}}]{Shorokhov2016}%
  \BibitemOpen
  \bibfield  {author} {\bibinfo {author} {\bibfnamefont {A.~S.}\ \bibnamefont
  {Shorokhov}}, \bibinfo {author} {\bibfnamefont {E.~V.}\ \bibnamefont
  {Melik-Gaykazyan}}, \bibinfo {author} {\bibfnamefont {D.~A.}\ \bibnamefont
  {Smirnova}}, \bibinfo {author} {\bibfnamefont {B.}~\bibnamefont {Hopkins}},
  \bibinfo {author} {\bibfnamefont {K.~E.}\ \bibnamefont {Chong}}, \bibinfo
  {author} {\bibfnamefont {D.~Y.}\ \bibnamefont {Choi}}, \bibinfo {author}
  {\bibfnamefont {M.~R.}\ \bibnamefont {Shcherbakov}}, \bibinfo {author}
  {\bibfnamefont {A.~E.}\ \bibnamefont {Miroshnichenko}}, \bibinfo {author}
  {\bibfnamefont {D.~N.}\ \bibnamefont {Neshev}}, \bibinfo {author}
  {\bibfnamefont {A.~A.}\ \bibnamefont {Fedyanin}}, \ and\ \bibinfo {author}
  {\bibfnamefont {Y.~S.}\ \bibnamefont {Kivshar}},\ }\href@noop {} {\bibfield
  {journal} {\bibinfo  {journal} {Nano Lett.}\ }\textbf {\bibinfo {volume}
  {16}},\ \bibinfo {pages} {4857} (\bibinfo {year} {2016})}\BibitemShut
  {NoStop}%
\bibitem [{\citenamefont {Shcherbakov}\ \emph
  {et~al.}(2015{\natexlab{b}})\citenamefont {Shcherbakov}, \citenamefont
  {Vabishchevich}, \citenamefont {Shorokhov}, \citenamefont {Chong},
  \citenamefont {Choi}, \citenamefont {Staude}, \citenamefont {Miroshnichenko},
  \citenamefont {Neshev}, \citenamefont {Fedyanin},\ and\ \citenamefont
  {Kivshar}}]{Shcherbakov2015b}%
  \BibitemOpen
  \bibfield  {author} {\bibinfo {author} {\bibfnamefont {M.~R.}\ \bibnamefont
  {Shcherbakov}}, \bibinfo {author} {\bibfnamefont {P.~P.}\ \bibnamefont
  {Vabishchevich}}, \bibinfo {author} {\bibfnamefont {A.~S.}\ \bibnamefont
  {Shorokhov}}, \bibinfo {author} {\bibfnamefont {K.~E.}\ \bibnamefont
  {Chong}}, \bibinfo {author} {\bibfnamefont {D.}~\bibnamefont {Choi}},
  \bibinfo {author} {\bibfnamefont {I.}~\bibnamefont {Staude}}, \bibinfo
  {author} {\bibfnamefont {A.~E.}\ \bibnamefont {Miroshnichenko}}, \bibinfo
  {author} {\bibfnamefont {D.~N.}\ \bibnamefont {Neshev}}, \bibinfo {author}
  {\bibfnamefont {A.~A.}\ \bibnamefont {Fedyanin}}, \ and\ \bibinfo {author}
  {\bibfnamefont {Y.~S.}\ \bibnamefont {Kivshar}},\ }\href@noop {} {\bibfield
  {journal} {\bibinfo  {journal} {Nano Lett.}\ }\textbf {\bibinfo {volume}
  {15}},\ \bibinfo {pages} {6985} (\bibinfo {year}
  {2015}{\natexlab{b}})}\BibitemShut {NoStop}%
\bibitem [{\citenamefont {Makarov}\ \emph {et~al.}(2015)\citenamefont
  {Makarov}, \citenamefont {Kudryashov}, \citenamefont {Mukhin}, \citenamefont
  {Mozharov}, \citenamefont {Milichko}, \citenamefont {Krasnok},\ and\
  \citenamefont {Belov}}]{Makarov2015}%
  \BibitemOpen
  \bibfield  {author} {\bibinfo {author} {\bibfnamefont {S.}~\bibnamefont
  {Makarov}}, \bibinfo {author} {\bibfnamefont {S.}~\bibnamefont {Kudryashov}},
  \bibinfo {author} {\bibfnamefont {I.}~\bibnamefont {Mukhin}}, \bibinfo
  {author} {\bibfnamefont {A.}~\bibnamefont {Mozharov}}, \bibinfo {author}
  {\bibfnamefont {V.}~\bibnamefont {Milichko}}, \bibinfo {author}
  {\bibfnamefont {A.}~\bibnamefont {Krasnok}}, \ and\ \bibinfo {author}
  {\bibfnamefont {P.}~\bibnamefont {Belov}},\ }\href@noop {} {\bibfield
  {journal} {\bibinfo  {journal} {Nano Lett.}\ }\textbf {\bibinfo {volume}
  {15}},\ \bibinfo {pages} {6187} (\bibinfo {year} {2015})}\BibitemShut
  {NoStop}%
\bibitem [{\citenamefont {Baranov}\ \emph {et~al.}(2016)\citenamefont
  {Baranov}, \citenamefont {Makarov}, \citenamefont {Milichko}, \citenamefont
  {Kudryashov}, \citenamefont {Krasnok},\ and\ \citenamefont
  {Belov}}]{Baranov2016}%
  \BibitemOpen
  \bibfield  {author} {\bibinfo {author} {\bibfnamefont {D.~G.}\ \bibnamefont
  {Baranov}}, \bibinfo {author} {\bibfnamefont {S.~V.}\ \bibnamefont
  {Makarov}}, \bibinfo {author} {\bibfnamefont {V.~A.}\ \bibnamefont
  {Milichko}}, \bibinfo {author} {\bibfnamefont {S.~I.}\ \bibnamefont
  {Kudryashov}}, \bibinfo {author} {\bibfnamefont {A.~E.}\ \bibnamefont
  {Krasnok}}, \ and\ \bibinfo {author} {\bibfnamefont {P.~A.}\ \bibnamefont
  {Belov}},\ }\href@noop {} {\bibfield  {journal} {\bibinfo  {journal} {ACS
  Photon.}\ }\textbf {\bibinfo {volume} {3}},\ \bibinfo {pages} {1546}
  (\bibinfo {year} {2016})}\BibitemShut {NoStop}%
\bibitem [{\citenamefont {Weismann}\ \emph {et~al.}(2015)\citenamefont
  {Weismann}, \citenamefont {Gallagher},\ and\ \citenamefont
  {Panoiu}}]{Weismann2015}%
  \BibitemOpen
  \bibfield  {author} {\bibinfo {author} {\bibfnamefont {M.}~\bibnamefont
  {Weismann}}, \bibinfo {author} {\bibfnamefont {D.~F.}\ \bibnamefont
  {Gallagher}}, \ and\ \bibinfo {author} {\bibfnamefont {N.~C.}\ \bibnamefont
  {Panoiu}},\ }\href@noop {} {\bibfield  {journal} {\bibinfo  {journal} {JOSA
  B}\ }\textbf {\bibinfo {volume} {32}},\ \bibinfo {pages} {523} (\bibinfo
  {year} {2015})}\BibitemShut {NoStop}%
\bibitem [{\citenamefont {Weismann}\ and\ \citenamefont
  {Panoiu}(2016)}]{Weismann2016}%
  \BibitemOpen
  \bibfield  {author} {\bibinfo {author} {\bibfnamefont {M.}~\bibnamefont
  {Weismann}}\ and\ \bibinfo {author} {\bibfnamefont {N.~C.}\ \bibnamefont
  {Panoiu}},\ }\href@noop {} {\bibfield  {journal} {\bibinfo  {journal} {Phys.
  Rev. B}\ }\textbf {\bibinfo {volume} {94}},\ \bibinfo {pages} {035435}
  (\bibinfo {year} {2016})}\BibitemShut {NoStop}%
\bibitem [{\citenamefont {Kruk}\ \emph {et~al.}(2015)\citenamefont {Kruk},
  \citenamefont {Weismann}, \citenamefont {Bykov}, \citenamefont {Mamonov},
  \citenamefont {Kolmychek}, \citenamefont {Murzina}, \citenamefont {Panoiu},
  \citenamefont {Neshev},\ and\ \citenamefont {Kivshar}}]{Kruk2015}%
  \BibitemOpen
  \bibfield  {author} {\bibinfo {author} {\bibfnamefont {S.}~\bibnamefont
  {Kruk}}, \bibinfo {author} {\bibfnamefont {M.}~\bibnamefont {Weismann}},
  \bibinfo {author} {\bibfnamefont {A.~Y.}\ \bibnamefont {Bykov}}, \bibinfo
  {author} {\bibfnamefont {E.~A.}\ \bibnamefont {Mamonov}}, \bibinfo {author}
  {\bibfnamefont {I.~A.}\ \bibnamefont {Kolmychek}}, \bibinfo {author}
  {\bibfnamefont {T.}~\bibnamefont {Murzina}}, \bibinfo {author} {\bibfnamefont
  {N.~C.}\ \bibnamefont {Panoiu}}, \bibinfo {author} {\bibfnamefont {D.~N.}\
  \bibnamefont {Neshev}}, \ and\ \bibinfo {author} {\bibfnamefont {Y.~S.}\
  \bibnamefont {Kivshar}},\ }\href@noop {} {\bibfield  {journal} {\bibinfo
  {journal} {ACS Photon.}\ }\textbf {\bibinfo {volume} {2}},\ \bibinfo {pages}
  {1007} (\bibinfo {year} {2015})}\BibitemShut {NoStop}%
\bibitem [{\citenamefont {Shen}(1984)}]{Shen1984}%
  \BibitemOpen
  \bibfield  {author} {\bibinfo {author} {\bibfnamefont {Y.~R.}\ \bibnamefont
  {Shen}},\ }\href@noop {} {\emph {\bibinfo {title} {The principles of
  nonlinear optics}}}\ (\bibinfo  {publisher} {Wiley-Interscience},\ \bibinfo
  {year} {1984})\BibitemShut {NoStop}%
\bibitem [{\citenamefont {Stegeman}\ and\ \citenamefont
  {Stegeman}(2012)}]{Stegeman2012}%
  \BibitemOpen
  \bibfield  {author} {\bibinfo {author} {\bibfnamefont {G.~I.}\ \bibnamefont
  {Stegeman}}\ and\ \bibinfo {author} {\bibfnamefont {R.~A.}\ \bibnamefont
  {Stegeman}},\ }\href@noop {} {\emph {\bibinfo {title} {Nonlinear Optics:
  Phenomena, Materials and Devices (Vol. 78)}}}\ (\bibinfo  {publisher} {John
  Wiley \& Sons},\ \bibinfo {year} {2012})\BibitemShut {NoStop}%
\bibitem [{\citenamefont {Slovick}\ \emph {et~al.}(2017)\citenamefont
  {Slovick}, \citenamefont {Zhou}, \citenamefont {Yu}, \citenamefont
  {Kravchenko}, \citenamefont {Briggs}, \citenamefont {Moitra}, \citenamefont
  {Krishnamurthy},\ and\ \citenamefont {Valentine}}]{Slovick2017}%
  \BibitemOpen
  \bibfield  {author} {\bibinfo {author} {\bibfnamefont {B.~A.}\ \bibnamefont
  {Slovick}}, \bibinfo {author} {\bibfnamefont {Y.}~\bibnamefont {Zhou}},
  \bibinfo {author} {\bibfnamefont {Z.~G.}\ \bibnamefont {Yu}}, \bibinfo
  {author} {\bibfnamefont {I.~I.}\ \bibnamefont {Kravchenko}}, \bibinfo
  {author} {\bibfnamefont {D.~P.}\ \bibnamefont {Briggs}}, \bibinfo {author}
  {\bibfnamefont {P.}~\bibnamefont {Moitra}}, \bibinfo {author} {\bibfnamefont
  {S.}~\bibnamefont {Krishnamurthy}}, \ and\ \bibinfo {author} {\bibfnamefont
  {J.}~\bibnamefont {Valentine}},\ }\href@noop {} {\bibfield  {journal}
  {\bibinfo  {journal} {Phil. Trans. R. Soc. A}\ }\textbf {\bibinfo {volume}
  {375}},\ \bibinfo {pages} {0072} (\bibinfo {year} {2017})}\BibitemShut
  {NoStop}%
\bibitem [{\citenamefont {Smith}\ \emph {et~al.}(2000)\citenamefont {Smith},
  \citenamefont {Padilla}, \citenamefont {Vier}, \citenamefont {Nemat-Nasser},\
  and\ \citenamefont {Schultz}}]{Smith2000}%
  \BibitemOpen
  \bibfield  {author} {\bibinfo {author} {\bibfnamefont {D.~R.}\ \bibnamefont
  {Smith}}, \bibinfo {author} {\bibfnamefont {W.~J.}\ \bibnamefont {Padilla}},
  \bibinfo {author} {\bibfnamefont {D.~C.}\ \bibnamefont {Vier}}, \bibinfo
  {author} {\bibfnamefont {S.~C.}\ \bibnamefont {Nemat-Nasser}}, \ and\
  \bibinfo {author} {\bibfnamefont {S.}~\bibnamefont {Schultz}},\ }\href@noop
  {} {\bibfield  {journal} {\bibinfo  {journal} {Phys. Rev. Lett.}\ }\textbf
  {\bibinfo {volume} {84}},\ \bibinfo {pages} {4184} (\bibinfo {year}
  {2000})}\BibitemShut {NoStop}%
\bibitem [{\citenamefont {Slovick}\ \emph {et~al.}(2013)\citenamefont
  {Slovick}, \citenamefont {Yu}, \citenamefont {Berding},\ and\ \citenamefont
  {Krishnamurthy}}]{Slovick2013}%
  \BibitemOpen
  \bibfield  {author} {\bibinfo {author} {\bibfnamefont {B.}~\bibnamefont
  {Slovick}}, \bibinfo {author} {\bibfnamefont {Z.~G.}\ \bibnamefont {Yu}},
  \bibinfo {author} {\bibfnamefont {M.}~\bibnamefont {Berding}}, \ and\
  \bibinfo {author} {\bibfnamefont {S.}~\bibnamefont {Krishnamurthy}},\
  }\href@noop {} {\bibfield  {journal} {\bibinfo  {journal} {Phys. Rev. B.}\
  }\textbf {\bibinfo {volume} {88}},\ \bibinfo {pages} {165116} (\bibinfo
  {year} {2013})}\BibitemShut {NoStop}%
\bibitem [{\citenamefont {Koschny}\ \emph {et~al.}(2003)\citenamefont
  {Koschny}, \citenamefont {Markos}, \citenamefont {Smith},\ and\ \citenamefont
  {Soukoulis}}]{Koschny2003}%
  \BibitemOpen
  \bibfield  {author} {\bibinfo {author} {\bibfnamefont {T.}~\bibnamefont
  {Koschny}}, \bibinfo {author} {\bibfnamefont {P.}~\bibnamefont {Markos}},
  \bibinfo {author} {\bibfnamefont {D.~R.}\ \bibnamefont {Smith}}, \ and\
  \bibinfo {author} {\bibfnamefont {C.~M.}\ \bibnamefont {Soukoulis}},\
  }\href@noop {} {\bibfield  {journal} {\bibinfo  {journal} {Phys. Rev. E}\
  }\textbf {\bibinfo {volume} {68}},\ \bibinfo {pages} {065602} (\bibinfo
  {year} {2003})}\BibitemShut {NoStop}%
\end{thebibliography}%

\end{document}